\documentclass[conference]{IEEEtran}
\IEEEoverridecommandlockouts
\usepackage{cite}
\usepackage{amsmath,amssymb,amsfonts}
\usepackage{algorithmic}
\usepackage{graphicx}
\usepackage{textcomp}
\usepackage{xcolor}
\usepackage{units}
\def\BibTeX{{\rm B\kern-.05em{\sc i\kern-.025em b}\kern-.08em
    T\kern-.1667em\lower.7ex\hbox{E}\kern-.125emX}}
\begin{document}

\title{External Electron Injection for the AWAKE Experiment}

\author{\IEEEauthorblockN{1\textsuperscript{st} Marlene Turner}
\IEEEauthorblockA{\textit{CERN} \\
Geneva, Switzerland \\
marlene.turner@cern.ch}
\and
\IEEEauthorblockN{2\textsuperscript{nd} Chiara Bracco}
\IEEEauthorblockA{\textit{CERN} \\
Geneva, Switzerland \\
}
\and
\IEEEauthorblockN{3\textsuperscript{th} Spencer Gessner}
\IEEEauthorblockA{\textit{CERN} \\
Geneva, Switzerland \\
}
\and
\IEEEauthorblockN{4\textsuperscript{th} Brennan Goddard}
\IEEEauthorblockA{\textit{CERN} \\
Geneva, Switzerland \\
}
\and
\IEEEauthorblockN{5\textsuperscript{th} Edda Gschwendtner}
\IEEEauthorblockA{\textit{CERN} \\
Geneva, Switzerland \\
}
\and
\IEEEauthorblockN{6\textsuperscript{th} Patric Muggli}
\IEEEauthorblockA{\textit{Max Planck Institute for Physics, CERN} \\
Munich, Germany \\
Geneva, Switzerland \\
}
\and
\IEEEauthorblockN{7\textsuperscript{th} Felipe Pe\~na Asmus}
\IEEEauthorblockA{\textit{Max Planck Institute for Physics} \\
Munich, Germany \\
}
\and
\IEEEauthorblockN{8\textsuperscript{rd} Francesco Velotti}
\IEEEauthorblockA{\textit{CERN} \\
Geneva, Switzerland \\
}
\and
\IEEEauthorblockN{9\textsuperscript{th} Livio Verra}
\IEEEauthorblockA{\textit{University of Milan, CERN} \\
Milan, Italy\\
Geneva, Switzerland \\
}
on behalf of the AWAKE Collaboration
}

\maketitle

\begin{abstract}
We summarize and explain the realization of witness particle injection into wakefields for the AWAKE experiment. In AWAKE, the plasma wakefields are driven by a self-modulating relativistic proton bunch. To demonstrate that these wakefields can accelerate charged particles, we inject a \unit[10-20]{MeV} electron bunch produced by a photo-injector. We summarize the experimental challenges of this injection process and present our plans for the near future.
\end{abstract}

\begin{IEEEkeywords}
AWAKE, Proton-Driven Plasma Wakefield Acceleration, Seeded-Self Modulation, External Electron Injection
\end{IEEEkeywords}

\section{Introduction}
Particle accelerators increase the energy (and velocity) of charged particles (traveling with velocities $v$ corresponds to a relativistic Lorentz factor $\gamma = 1/\sqrt{(1-v^2/c^2)}$, where $c$ is the speed of light). Every accelerator consists of at least a particle source (injector) and an accelerating section. In electron accelerators that use radio-frequency cavities, the source is often a photo-injector or a thermionic electron source. The injected particles increase their energy if they are captured in the accelerating (and focusing) phase of the radio-frequency fields.

\begin{figure*}[ht!]
\begin{center}
\includegraphics[width=0.8\textwidth]{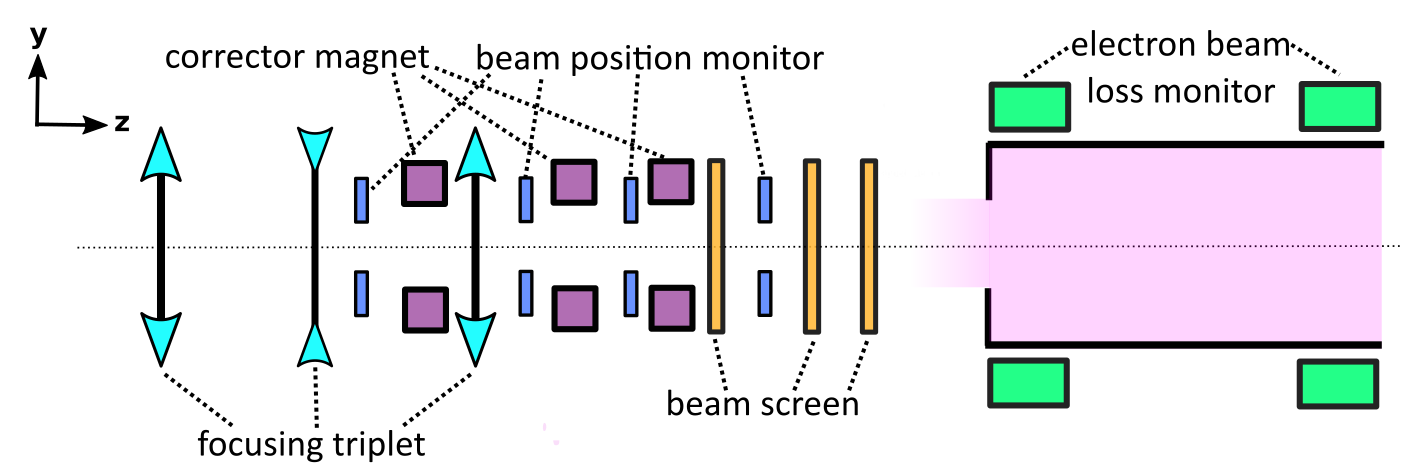}
\label{fig:schematic}
\caption{Schematic drawing of the optics and diagnostics setup of the electron beamline upstream and along the plasma. Drawing not to scale.}
\end{center}
\end{figure*}

In plasma-based accelerators, a traveling plasma electron density modulation sustains transverse and longitudinal fields. In linear theory wakefields are sinusoidal with a periodicity of the plasma electron wavelength $\lambda_{pe}$. Transverse and longitudinal wakefields are $\lambda_{pe}/4$ out of phase. Longitudinal wakefields can accelerate or decelerate charged particles; transverse wakefields either focus or defocus them. Various particle injection schemes have been developed, for example: self- \cite{selfinj}, external-, trojan horse- \cite{trojhorse}, plasma down ramp- \cite{downramp} and ionization injection \cite{ionization}. The selection of the injection scheme depends on the experimental setup and parameters such as the plasma electron density, drive bunch phase velocity, etc.

As described for example in \cite{trapping1,trapping2} (1D theory), a charged particle (with a relativistic factor $\gamma$) can be longitudinally trapped by a plasma wave with a given phase velocity $\beta_{ph}=\sqrt{1-1/\gamma_{ph}^2}$, if its energy is larger than $\gamma_{min}mc^2$ and smaller than $\gamma_{max}mc^2$, where $m$ the particles rest-mass. The minimum and maximum $\gamma$ can be calculated according to: 
\begin{equation}
\gamma_{max,min}= \gamma_{ph}(1+\gamma_{ph}\Delta \phi) \pm \gamma_{ph}\beta_{ph}[(1+\gamma_{ph}\Delta\phi)^2-1]^{1/2}
\label{eq:capture}
\end{equation}
where $\Delta \phi = 2 \beta_{ph}[(1+E_{max}^2/2)^2-1]^{1/2}$ \cite{trapping2}, $E_{max}=E/E_0$, $E$ the wakefield amplitude, $E_0=mc\omega_{pe}/e$ the cold plasma wave-breaking field with $\omega_{pe}$ the plasma electron frequency and $e$ the electron charge.

Charged particles increase their longitudinal energy as long as they are located in the accelerating phase of the wakefield. The phase velocity of the plasma wave $\beta_{ph}$ is approximately the velocity of the driver $\beta_{p}$. Often, experiments are designed such that the wakefields phase velocity does not change significantly over the length of the plasma. On the contrary, witness particles accelerate (often from very low energies) along the plasma and thus their velocity changes. When the phase slippage between the two bunches is larger than $\sim\lambda_{pe}/4$ witness particles either enter the focusing, decelerating phase of the wakefield (they fall behind); or the defocusing, accelerating phase (they outrun the wave).

There can be other reasons for witness particles to dephase in the wakefields. For example: when the plasma density along the propagation axis is non-uniform, the plasma electron wavelength $\lambda_{pe}$ changes; when the drive bunch distribution changes along the plasma, the wakefield phase (and velocity) also evolves.

In this article, we discuss electron injection into a plasma wave that is driven by a self-modulating proton bunch in the context of the AWAKE experiment. In practice, the injection faces two major challenges: 1) as the drive bunch is self-modulating over the first few meters of plasma, its distribution is evolving; 2) there is a plasma density ramp at the plasma entrance. Consequently, the wakefield's velocity is evolving in the ramp and along the first few meters of plasma \cite{schroeder,einjlotov,einjalexey}. Additionally the initial seed wakefields driven by the unmodulated proton bunch are mostly defocusing for electrons.

To avoid defocusing of the witness electrons close to the plasma entrance we aim to inject electrons at a certain longitudinal location inside the plasma, by crossing the electron with the proton bunch trajectory\cite{einjlotov} (i.e. also the wakefields). This injection geometry avoids the effect of the ramp and places the electrons at a location where they can be trapped. We discuss the realization of this injection scheme in the AWAKE experiment as well as the experimental challenges.

\subsection{The AWAKE experiment}
The Advanced Wakefield Experiment (AWAKE) \cite{AWAKE} accelerates electrons with plasma wakefields driven by a relativistic proton bunch. To reach GV/m field amplitudes, the experiment operates at plasma electron densities in the range of \unit[$(1-10)\times10^{14}$]{cm$^{-3}$} (plasma electron wavelength $\lambda_{pe}\simeq \unit[1-3]{mm}$). Since the \unit[400]{GeV/c} proton drive bunch ($\gamma=427$) --as delivered by the CERN Super Proton Synchrotron-- has a rms bunch length on the order of \unit[$\sim$6-12]{cm}, it is much too long to effectively drive wakefields at these plasma densities. Thus, the experiment relies on the seeded self-modulation \cite{SSM} to modulate the proton bunch density at the plasma electron wavelength. The bunch train then resonantly excites a plasma wave with hundreds of MV/m field amplitudes. 

The plasma is created by a \unit[120]{fs}, \unit[$<$450]{mJ} laser pulse that ionizes the outermost electron of each rubidium atom in a \unit[10]{m} long rubidium vapor source \cite{vaporsource}. There is no window that allows a proton, laser and electron bunch to enter the vapor source without significant distortion (of at least one of them). Thus they enter the source through an aperture with a diameter of \unit[10]{mm}. Rubidium flowing out of the apertures condensates on the cold walls of an expansion volume and creates a vapor density ramp with a length that is on the order of the opening aperture \cite{plasma}.

\section{Experimental Realization of External Electron Injection}

Once the self-modulation process saturates, the proton bunch resonantly drives wakefields with $\gamma_{ph}\simeq427$. The seed fields are on the order of \unit[10]{MV/m}. To obtain a first estimate for the initial witness particle energy necessary to be trapped, we use Eq. \ref{eq:capture}. It shows that to longitudinally trap particles at the plasma entrance with the seed fields, their Lorentz factor must be between $\gamma \simeq26$ and $6900$. To trap particles \unit[$\approx 1-2$]{m} into the plasma, where wakefield amplitudes are expected to reach at least \unit[100]{MV/m} it must be between $\gamma \simeq 6$ and $30000$. We note that this is just a rough estimate since, as mentioned before, the phase velocity and amplitude of the wakefield vary due to the development of the SSM.

Thus in the AWAKE facility we produce a \unit[$\sim$5]{MeV}, \unit[$\sim$8]{ps} (\unit[$\sim3$]{mm}) long electron bunch with a photo-injector and accelerate it to \unit[10-20]{MeV} ($\gamma\simeq 19-39$) in a \unit[1]{m} long booster structure (detailed description in \cite{esource}). The beam is then transported to the entrance of the plasma (detailed description in \cite{ebeam1,ebeam2}) and focused with the final triplet (see Fig. \ref{fig:schematic}). 

We note here that the length of the electron bunch is on the order of $\lambda_{pe}$. This long length was chosen for initial experiments to avoid the need for precise timing between the electron bunch and the wakefield. It guarantees some charge capture for each (properly aligned) injection event. To capture and accelerate electrons, we overlap in space the electron bunch trajectory with the plasma wakefields, within the self-modulating proton bunch.

\begin{figure}[htb]
\begin{center}
\includegraphics[width=1\columnwidth]{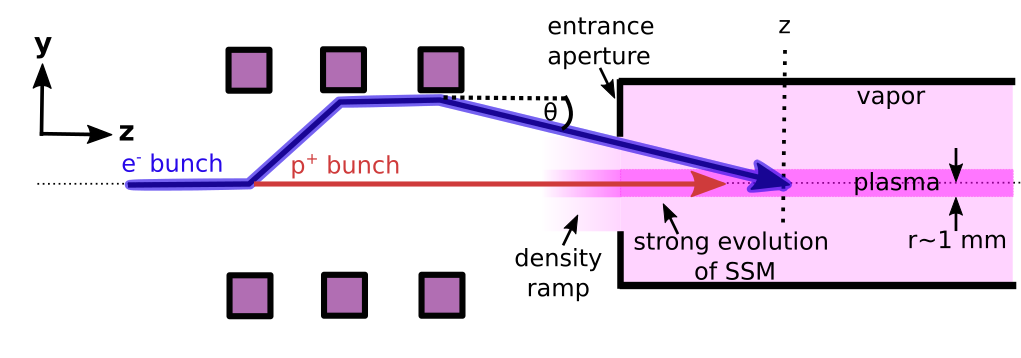}
\label{fig:schematic_inj}
\caption{Schematic drawing of the electron injection scheme in the vertical plane (only corrector magnets are shown, see Fig. \ref{fig:schematic}). The proton bunch propagates along the horizontal axis (horizontal black dotted line).}
\end{center}
\end{figure}

To inject the electrons sideways, we use the first two corrector magnets in Fig. \ref{fig:schematic} and \ref{fig:schematic_inj} to create a vertical parallel offset (maximum vertical offset \unit[$\Delta y \sim$15]{mm}) of the electron beam with respect to the proton beam trajectory. We use the third corrector magnet to set the desired injection angle towards the proton bunch trajectory (see Figure \ref{fig:schematic_inj}). The electrons propagate in vacuum or vapor (once they enter the vapor source) and do not experience strong fields until they enter the plasma (plasma radius \unit[$\sim1$]{mm}) and the wakefields. 

Due to the \unit[r=5]{mm} entrance aperture of the vapor source, the maximum injection angle is given by $\theta_{max}$ = min($ \Delta y/(L+3.4)$, \unit[5]{mm}/$L$), where $L$ is the distance between the electron-wakefield crossing point and the plasma entrance (in meters). Horizontally, we align the electron beam to the proton beam trajectory.

\subsection{Beam Instrumentation}

To monitor the electron bunch upstream the plasma, we use beam position monitors (BPM), scintillating beam screens and electron beam loss monitors (eBLMs) \cite{beaminstr} (see Fig. \ref{fig:schematic}). The beam screens measure the position and transverse electron bunch distribution even in the presence of the proton bunch. Since the electrons experience large scattering angles in the screen material, we can only measure the electron bunch with one screen at a time. The BPMs monitor the position of the charge center of the electron bunch along the transport line. We must use screens displaying the electron and proton bunch in order to find an absolute reference trajectory on the various BPMs. We use screens as well as BPMs to align the electron bunch with respect to the proton bunch trajectory (and thus the wakefields). The eBLMs detect signals of secondary particles that are produced when electrons (and protons) interact with the vapor source material (e.g. aperture plate). We have installed eight of them around the entrance and along the vapor source. 

\section{Alignment Challenges}
\label{sec:challenges}
We aim to cross the electron bunch (focused at the crossing point) with the proton bunch and the plasma wakefields at a defined longitudinal position $z$ (see Fig. \ref{fig:schematic_inj}) inside the plasma. Additionally, we want to control the electron injection angle $\theta$. Precise alignment is required as: 1) the transverse extent of the plasma wakefields is small (on the order of a few plasma skin depths $c/\omega_{pe}$, i.e. \unit[$\sim$0.5-0.1]{mm} for our plasma electron densities); 2) the nominal electron bunch rms size at the focus is small ($\sigma_r\sim\unit[0.2]{mm}$) 3) there is no diagnostics for bunch crossing inside the plasma. In the following subsections we discuss various challenges regarding electron injection.

\subsection{Earth's Magnetic Field}
The energy of the incoming electron bunch is low ($\unit[10-20]{MeV}$). The earth's magnetic field $B$ ($B_x$\unit[$\sim$0.2]{Gauss}, $B_y$\unit[$\sim$0.4]{Gauss}, corresponding to a Larmor radius $r_L = \beta \gamma m c /eB$ of $\sim 1.5$ and \unit[3]{km}, respectively) leads to a significant electron deflection angle even over short distances (the effect of the earth's magnetic field on the \unit[400]{GeV/c} proton bunch is negligible).

To quantify this effect experimentally, we centered the proton bunch on the plasma entrance aperture and aligned the electron bunch onto the proton trajectory at the first two beam screens (see Fig. \ref{fig:schematic}). We then used the last corrector magnet in Fig. \ref{fig:schematic_inj} to scan the electron bunch position horizontally and vertically across the entrance aperture, while recording the eBLM loss signal. The asymmetry in the eBLMs signal versus corrector magnet setting yields the trajectory angle due to the earth's magnetic field (constant in amplitude). We compensate the angle with the last corrector magnets to reach true electron-proton beam alignment at the plasma entrance. This trajectory can then be used as a reference for injection.  

\subsection{Electron-- Proton Bunch Interaction}
The electron and proton bunch represent opposite currents that attract each other ($\propto \gamma m$). Since in our experiment the electron bunch can be affected by the proton bunch, we experimentally studied their radial attraction. We set the two beam trajectories to be parallel (over \unit[$\sim$3-15]{m}) with a distance of \unit[$~1-10$]{mm} after the last quadrupole magnet (see Fig. \ref{fig:schematic}). We measured on screens upstream and downstream the plasma the position of the electron bunch without and with presence of protons. No measurable position change was observed. The effect is thus negligible in these experiments.

\subsection{Layout constraints}
To enter the vapor source, the electron bunch has to pass through the entrance aperture. When the focal point of the beam is set to be inside the plasma we lose electrons as the beam size at this aperture becomes larger than \unit[$\sigma_r\sim1-2$]{mm} (in our injection scheme we use \unit[0-4]{mm} offsets at the entrance aperture, as shown in Fig. \ref{fig:schematic_inj}).

Initial experiments were performed with a relatively large emittance electron beam (normalized emittance \unit[$\epsilon_N=$14]{mm mrad} instead of the baseline \unit[$\epsilon_N=$2]{mm mrad} value). In this case, placing the electron waist some distance in the plasma, makes the radial beam size large at the vapor source aperture location. Measurements with the eBLMs show that with the beam envelope size imposed by the magnetic optics and when focused \unit[5]{m} inside the source, only \unit[$\sim$30]{\%} of the charge enters the plasma  (on-axis alignment). To avoid this beam loss for first acceleration experiments (see Sec. \ref{sec:results}), we focused the beam at the plasma entrance. 

\subsection{Diagnostics limitations}
The BPMs are not able to measure the electron beam position in the presence of the proton bunch. Their signal is overwhelmed by that of the large proton bunch population (\unit[3$\times10^{11}$]{protons/bunch} versus \unit[0.6-3.6$\times10^{9}$]{electrons/bunch}). For the acceleration experiments we must also remove beam screens so that the electrons and the ionizing laser pulse can enter the vapor source. To have an idea of where and how we inject electrons into the wakefields, we developed two techniques: 1) we analyzed the measurements from electron beam position monitors (\unit[10]{Hz}) in between proton events (\unit[$\sim 0.3$]{Hz}) and computed an average electron beam trajectory; 2) we used the upstream BPMs (not affected by the proton bunch), the energy jitter (as computed from strategically placed BPMs) and the fitted betatronic oscillations to predict the trajectory variations around the reference\cite{reconstruction}. This allows us to reconstruct the electron-proton crossing point for each measured event.

\subsection{Technical issues}
We observed that changing the current of the magnets in: a) the electron spectrometer downstream the plasma and b) the proton transport line upstream the plasma changes the trajectory of the incoming electron bunch. Consequently we align the beam with those magnets at their nominal current values and re-steer the electron bunch after measurable changes.

Due to the strong focusing of the last triplet (see Fig. \ref{fig:schematic}) and the long distance between the plasma-electron crossing point and the beam diagnostic (\unit[$\sim$3-5]{m}), the transverse electron bunch size at the beam screens is large (compared to the transverse proton bunch size). This makes the determination of the bunch position and its trajectory, and thus the spatial injection alignment, challenging.

\section{First Electron Acceleration Experiments}
\label{sec:results}

The AWAKE collaboration started electron acceleration experiments in May 2018. The incoming electron bunch had a transverse emittance of \unit[$\sim14$]{mm mrad} with a charge of \unit[$650$]{pC}, out of which \unit[$\sim 70-80$]{\%} were transported through the vapor source aperture. In those experiments, we accelerated electrons from \unit[$\sim18$]{MeV} up to \unit[2]{GeV}\cite{AWAKEAcceleration}. 

\begin{figure}[htb]
\begin{center}
\includegraphics[width=1\columnwidth]{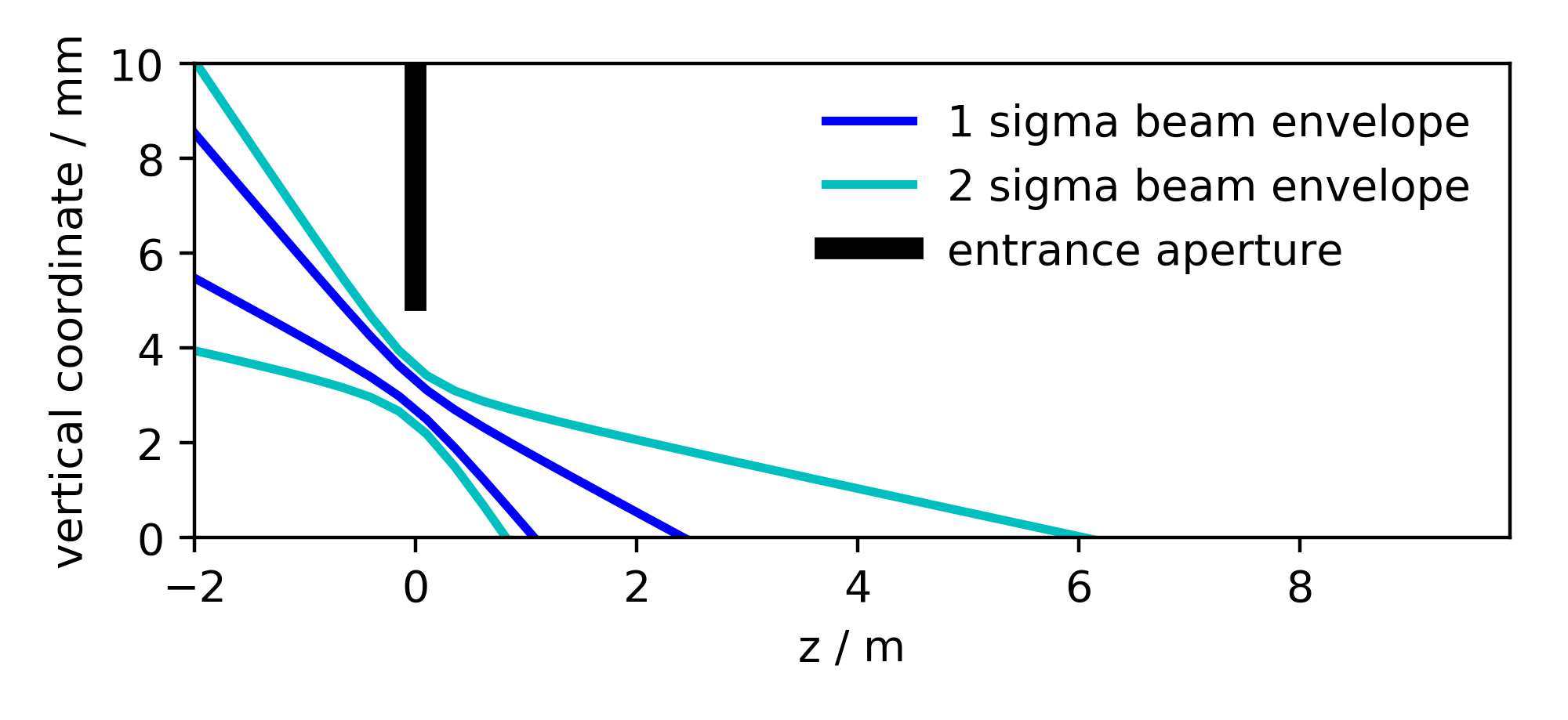}
\label{fig:inj}
\caption{Theoretical electron beam envelope during first AWAKE injection experiments.}
\end{center}
\end{figure}

Due to challenges mentioned in Section \ref{sec:challenges}, we focused the electron bunch at the entrance of the plasma. At that location, the focal point was set \unit[$\sim$3-4]{mm} above the proton trajectory. The electron bunch trajectory was set to be at an angle of \unit[$\sim$2]{mrad} with respect to the proton bunch trajectory. Thus the two trajectories crossed \unit[$\sim2$]{m} past the vapor source aperture. The theoretical envelope of the electron beam during injection is shown in Fig. \ref{fig:inj}. We note that rather than injecting the electron bunch at a defined location, we 'sprayed' them into the plasma and the wakefields. 

\section{Towards Controlled Electron Injection}
While we successfully injected and accelerated a small amount of electrons (\unit[$\sim$0.3]{pC})\cite{AWAKEAcceleration}, we did not know: a) at which location along the plasma those electrons crossed the wakefields; b) what their injection angle was. Simulation results suggest that electron trapping and acceleration is very sensitive to both of those parameters \cite{einjlotov,einjalexey}.
 
To improve the situation for the next experiments, we:
\begin{itemize}
\item have decreased the transverse emittance of the incoming electron bunch at a bunch charge of \unit[$\sim$100]{pC} (currently \unit[3-4]{mm mrad}) \cite{emittance};
\item have worked on different optics, in which the beam size at the entrance aperture is smaller (at the expense of a larger focal point size)\cite{optics}.
\end{itemize} 
This means that we are able to: 1) focus the electron bunch at different locations inside the plasma; 2) cross the wakefields with different incoming angles, without losing a significant amount of charge on the vapor source aperture.

This controlled electron injection will allow us to:
\begin{itemize}
\item understand from which point on the wakefield's phase is stable enough to capture and transport electrons;
\item measure the energy of the accelerated electrons as a function of injection position along the plasma, to map the accelerating field gradient along the plasma;
\item understand how charge capture depends on the injection angles. Simulation results predict strong dependencies.
\end{itemize}
Measurements are scheduled for autumn 2018.

\section{Summary and Conclusions}

We summarized the realization of external electron injection for the first run of the AWAKE experiment. Wakefield phase evolution due to density ramp and development of the SSM may cause loss of the electron witness particles in the beginning of the plasma. Thus we displace vertically the electron trajectory and inject the electron bunch obliquely into the wakefield, where their phase is not evolving significantly anymore. We additionally listed challenges we encountered, studies we performed and our plans for the near future. The first experimental injection experiments using the scheme described here led to acceleration of electrons up to \unit[2]{GeV}\cite{AWAKEAcceleration}.



\begin{thebibliography}{00}
\bibitem{selfinj} T. Tajima and J.M. Dawnson, ``Laser Electron Accelerator,'' Phys. Rev. Lett., 43, 267, 1979.
\bibitem{trojhorse} B. Hidding et al., ``Ultracold Electron Bunch Generation via Plasma Photocathode Emission and Acceleration in a Beam-Driven Plasma Blowout,'' Phys. Rev. Lett., 108, 035001, 2012.
\bibitem{downramp} M. Hansson et al., ``Down-ramp injection and independently controlled acceleration of electrons
in a tailored laser wakefield accelerator,'' Phys. Rev. Lett., 18, 071303, 2015.
\bibitem{ionization} E. Oz et al., ``Ionization-Induced Electron Trapping in Ultrarelativistic Plasma Wakes,'' Phys. Rev. Lett., 98, 084801,  2007.
\bibitem{trapping1} C. Joshi et al., ``Forward Raman Instability and Electron Acceleration,'' Phys. Rev. Lett., vol. 47, number 18, 1981.
\bibitem{trapping2} E. Esarey and M. Pilloff, ``Trapping and Acceleration in Nonlinear Plasma Waves,'' Phys. of Plasmas, 1432, 1995.
\bibitem{schroeder} C.B. Schroeder et al., ``Growth and Phase Velocity of Self-Modulated Beam-Driven Plasma Waves,''  Phys. Rev. Lett., vol. 107, 145002,  2011.
\bibitem{einjlotov} K.V. Lotov et al., ``Electron trapping and acceleration by the plasma wakefield of a self-modulating proton beam,''  Proceedings of IPAC 2014.
\bibitem{einjalexey} A. Petrenko et al., ``Electron Injection Studies for the AWAKE Experiment at CERN,''  Proceedings of IPAC 2014.
\bibitem{AWAKE} E. Gschwendtner et al., ``AWAKE, The Advanced Proton Driven Plasma Wakefield Acceleration Experiment at CERN,''  NIM A Volume 829, 1  2016.
\bibitem{SSM} P. Muggli and the AWAKE Collaboration, ``AWAKE readiness for the study of the seeded self-modulation of a 400 GeV proton bunch,''  
Plasma Physics and Controlled Fusion, 2017.
\bibitem{vaporsource} E. Oz, P. Muggli, ``A novel Rb vapor plasma source for plasma wakefield accelerators,'' NIM A 740 2014.
\bibitem{plasma} G. Plyushchev et al., ``A Rubidium Vapor Source for a Plasma Source for AWAKE,''  Journal of Physics D: Applied Physics, 51(2), 025203 (2017).
\bibitem{esource} K. Pepitone et al., ``The electron accelerators for the AWAKE experiment at CERN Baseline and Future Developments,''  2018.
\bibitem{ebeam1} J.S. Schmidt et al., ``The AWAKE Electron Primary Beamline,''  Proceedings of IPAC 2015.
\bibitem{ebeam2} J.S. Schmidt et al., ``Status of the proton and electron transfer lines for the AWAKE Experiment at CERN,''  NIM A 892 2016.
\bibitem{beaminstr} S. Mazzoni et al., ``Beam Instrumentation Developments for the Advanced 
Proton Driven Plasma Wakefield Experiment at CERN ,'' Proceedings of IPAC 2017.
\bibitem{reconstruction} F. Velotti, CERN, private communication.
\bibitem{AWAKEAcceleration} E. Adli and the AWAKE Collaboration, ``Acceleration of electrons in the plasma wakefield of a proton bunch,'' Nature 561,363 2018.
\bibitem{emittance} S. Doebert, CERN, private communication.
\bibitem{optics} C. Bracco, CERN, private communication.
\end{thebibliography}
\end{document}